\documentclass[11pt]{article}

\usepackage{amsmath,amssymb}
\usepackage{graphicx,color}
\newcommand{\fund}
{\setlength{\unitlength}{0.45pt}
\begin{picture}(10,10)
\put(0,0){\line(1,0){10}}
\put(0,10){\line(1,0){10}}
\put(0,0){\line(0,1){10}}
\put(10,0){\line(0,1){10}}
\end{picture}}

\newcommand{\fundl}
{\setlength{\unitlength}{0.6pt}
\begin{picture}(10,10)
\put(0,0){\line(1,0){10}}
\put(0,10){\line(1,0){10}}
\put(0,0){\line(0,1){10}}
\put(10,0){\line(0,1){10}}
\end{picture}}

\newcommand{\anti}
{\setlength{\unitlength}{0.45pt}
{\begin{picture}(10,20)
\put(0,0){\line(0,1){20}}
\put(0,20){\line(1,0){10}}
\put(0,10){\line(1,0){10}}
\put(0,0){\line(1,0){10}}
\put(10,0){\line(0,1){20}}
\end{picture}}}

\newcommand{\antis}
{\setlength{\unitlength}{0.45pt}
{\begin{picture}(10,20)
\put(0,-5){\line(0,1){20}}
\put(0,15){\line(1,0){10}}
\put(0,5){\line(1,0){10}}
\put(0,-5){\line(1,0){10}}
\put(10,-5){\line(0,1){20}}
\end{picture}}}

\newcommand{\antin}
{\setlength{\unitlength}{0.4pt}
{\begin{picture}(10,20)
\put(0,2){\line(0,1){23}}
\put(0,25){\line(1,0){10}}
\put(0,15){\line(1,0){10}}
\put(0,15){\line(1,0){10}}
\put(10,2){\line(0,1){23}}
\put(2,-6){.}
\put(2,-2){.}
\put(0,5){\line(1,0){10}}
\put(10,-20){\line(0,1){13}}
\put(0,-20){\line(0,1){13}}
\put(0,-10){\line(1,0){10}}
\put(0,-20){\line(1,0){10}}
\end{picture}}}

\numberwithin{equation}{section}
\begin{document}

\quad 
\vspace{-3.5cm}

\begin{flushright}
\parbox{3cm}
{
{\bf May 2008}\hfill \\
UT-08-14 \hfill \\
 }
\end{flushright}

\vspace*{0.5cm}

\begin{center}
\Large\bf 
Flop Invariance of Refined Topological Vertex\\ and Link Homologies
\end{center}
\vspace*{0.7cm}
\centerline{\large 
Masato Taki
}
\begin{center}
\emph{Department of Physics, Faculty of Science, University of Tokyo,\\
Bunkyo-ku, Tokyo 113-0033, Japan.} \\
\vspace*{0.5cm}
{\tt tachyon@hep-th.phys.s.u-tokyo.ac.jp}
\end{center}

\vspace*{0.7cm}

\centerline{\bf Abstract} 

\vspace*{0.5cm}

It has been proposed recently that the topological A-model string theory on local toric Calabi-Yau manifolds has a two parameter extension.  Amplitudes of the two parameter topological strings can be computed using a diagrammatic method called the refined topological vertex. In this paper we study properties of the refined amplitudes under the flop transition of toric Calabi-Yau three-folds. We also discuss that the slicing invariance and the flop transition imply a simple formula for the homological $sl(N)$ invariants of the Hopf link. The new expression for the invariants gives a simple refinement of the Hopf link invariant of Chern-Simons theory.

\vspace*{1.0cm}

\vfill

\thispagestyle{empty}
\setcounter{page}{0}

\newpage

\section{Introduction}
Topological string theory is a special class of two-dimentional topological sigma model coupled to two-dimentional gravity. 
One of the importances of topological strings is that they provide many insights into physics and mathematics \cite{review}. 

There exist two types of topological strings according to the twist procedures in ${\cal N}=(2,2)$ sigma model which are called A-model and B-model respectively. In this articles we study A-model topological strings for local Calabi-Yau manifolds. In general topological strings partition function have the following form
\begin{equation}
Z (q=e^{-\hbar},t_i)=  \exp{{\cal F}(\hbar ,t_i )}  = \exp{\sum\limits_{g = 0}^\infty  {\hbar^{2g - 2} {\cal F}_g ( t_i )} },
\end{equation}
where $\hbar$ is the topological string coupling constant and $t_i$ are K\"{a}hler parameters of the geometry. It is known that A-model partition functions for the local toric Calabi-Yau threefolds can be solved by the topological vertex formalism \cite{Aganagic:2003db}. Using these partition functions we can count BPS states of effective theories of Type IIA superstrings on a toric Calabi-Yau manifold in the presence of the self-dual graviphoton background $F_{12}=F_{34}=\hbar$.

Recently the refined topological vertex formalism has been proposed via the melting crystal picture of A-model \cite{Iqbal:2007ii}. From the target-space perspective the refined vertex captures information about the Lefshetz action on the moduli space of the BPS states. Therefore it enable us to count the BPS particles in the non self-dual background $F_{12}=\epsilon_1$, $F_{34}=\epsilon_2$ and reproduce the K-theoretic Nekrasov formulas from string theory. Corresponding to the background, the refined vertex has two parameters $t=e^{-\epsilon_1}$, $q=e^{-\epsilon_2}$.

In the first half of this article we study transformation properties of refined partition functions under the flop transition of toric Calabi-Yau manifolds. The flop invariance of the topological vertex has been studied in \cite{Iqbal:2004ne}\cite{Konishi:2006ev}. So we generalize their arguments to the refined vertex using the free fermion techniques \cite{Okounkov:2003sp}\cite{Eguchi:2003it}.

In the latter half of the paper we apply our results to the homological link invariants of Hopf link. By assuming that the slicing invariance  \cite{Iqbal:2007ii}\cite{Iqbal:2008ra} is satisfied we find a simple formula for the superpolynomial of  \cite{Gukov:2007tf}. Using this result, one can save computational costs because there exists no need to sum up the Macdonald functions. 

This paper is organized as follows.
In section 2, we study the invariance of the refined topological vertex under the flop transition. The application of the flop to the homological link invariant is studied and the formula for the superpolynomial is proposed in section 3.
Conclusions are found in section 4. In appendix A, we give some useful formulas for the Schur functions.

%%%%%%%%%%%%%%%%%%%%%%%%%%%%%%%%%%%%%%%%%
\section{Flop Transition of Reined Vertex}
\subsection{Calculation of Subdiagrams} 

The topological vertex formalism is a cut-and-paste method to compute A-model partition functions for toric Calabi-Yau manifolds. The method consists of some rules. Let us consider A-model on a toric Calabi-Yau. First we divide the web diagram of the Calabi-Yau into trivalent vertices and propagator lines. On the edges of these parts we associate Young diagrams as an analogue of momentums for the Feynman rules. Then we associate the vertex functions $C_{\lambda, \mu,\nu}$ and the propagators $(-Q)^{|\lambda|}\delta_{\lambda,\lambda^{\prime}}$ for these parts. Here $Q=e^{-t}$ is a K\"{a}hler parameter of a corresponding $\mathbb{P}^1$. Then inserting some factors\footnote[1]{There is no framing factor for the conifold.} coming from the framing dependence we glue these parts into one partition function $Z(q,Q)$. The gluing procedure is done by summing over the Young diagrams.
  
\begin{figure}[h]
\begin{center}
\includegraphics[width=3cm,clip]{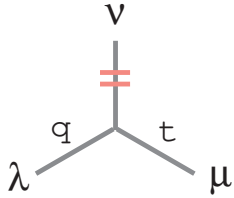}\\
\end{center}
\caption{The refined topological vertex $\mathop C\nolimits_{\lambda \mu \nu } (t,q)$}
\label{refined}
\end{figure}
The refined vertex function for the sub-diagram Fig.\ref{refined} which is proposed in \cite{Iqbal:2007ii} is such that
\begin{align}
C_{\lambda \mu \nu } (t,q) ={\left( {\frac{q}{t}} \right)}^{\frac{{\left\| \mu  \right\|^2  + \left\| \nu  \right\|^2 }}{2}} t^{\frac{{{\kappa}_\mu  }}{2}} P_{{\nu}^t } (\mathop t\nolimits^{ - \rho } ;q,t)\sum\limits_\eta  {{\left( {\frac{q}{t}} \right)}^{\frac{{\left| \eta  \right| + \left| \lambda  \right| - \left| \mu  \right|}}{2}} s_{\lambda ^t /\eta } (t^{ - \rho } q^{ - \nu } ) s_{\mu /\eta } (t^{ - \nu^t } q^{ - \rho } )} 
\end{align}
where $P_{{\nu}^t } (\mathop t\nolimits^{ - \rho } ;q,t)$ is a specialization of the Macdonald function \cite{Awata:2005fa}
\begin{align}
P_{{\nu}^t } (\mathop t\nolimits^{ - \rho } ;q,t)=t^{\frac{1}{2}||\nu||^2}\tilde{Z}_{\nu}(t,q),
\quad
\tilde{Z}_{\mu}(t,q)=\prod_{(i,j)\in\nu}(1-t^{\nu_j^t-i+1}q^{\nu_i-j})^{-1}.
\end{align}
The preferred direction with representation $\nu$ is indicated by short red lines in Fig.\ref{refined}.

\begin{figure}[htbp]
\begin{center}
\includegraphics[width=9cm,clip]{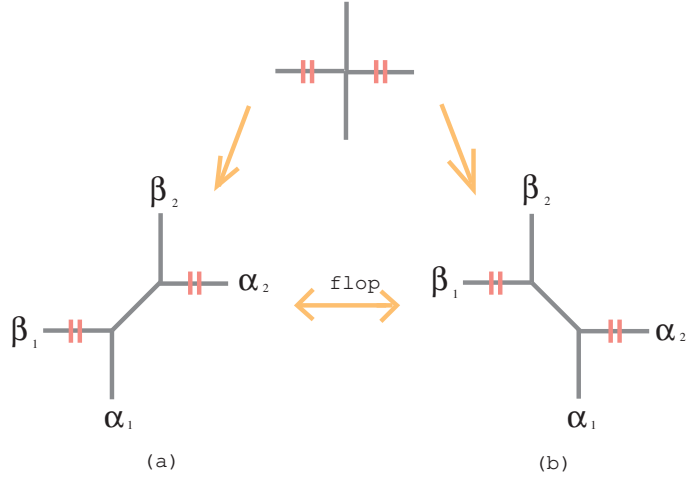}\\
\end{center}
\caption{The two resolved conifolds (a) and (b) are related via the flop transition.}
\label{floptr}
\end{figure}

In this section we calculate topological string maplitudes for Fig.\ref{floptr} in order to show that the refined patririon functions possess the invariances under the flop transitions. The flop invariance of the topological vertex has been studied in \cite{Iqbal:2004ne}\cite{Konishi:2006ev}. So we generalize their arguments for the refined topological vertex. By the rules of the topological vertex, the partition function of Fig.2(a) is given by
%%%%%%%%%%%%%%%%%%%%%%%%%%%%%%%%%%%%%%%%%%%%%%%%%%%%%%%%%%%%%
\begin{align}
Z^{(\textrm{a})} (Q,t,q) &= \sum_\alpha  {{( - Q)}^{\left| \alpha  \right|} C_{\alpha _1 \alpha \beta_1 } (t,q) C_{ \beta_2 \alpha^t \alpha_2} (q,t)} \nonumber\\
 &= \sum_{\alpha,\tau,\sigma}  { {( - Q)}^{\left| \alpha  \right|} 
{\left( {\frac{q}{t}} \right)}^{\frac{1}{2}( {\left\| \alpha  \right\|}^2  + 
{\left\| {\beta_1 } \right\|}^2 )} t^{\frac{1}{2} \kappa_\alpha  } 
P_{\beta_1 ^t } (t^{-\rho};q,t)} 
%\sum_\tau  
{ {\left( {\frac{q}{t}} \right)}^{\frac{1}{2}(\left| \tau  \right| + \left| { \alpha _1 } \right| - \left| \alpha  \right|)}  s_{ \alpha_1 ^t /\tau } ( t^{ - \rho } q^{ -  \beta_1 } ) s_{\alpha /\tau } (q^{ - \rho }  t^{ -  \beta_1 ^t } )} \nonumber\\
&\times {\left( {\frac{t}{q}} \right)}^{\frac{1}{2}( {|| {\alpha^t } ||}^2
+ {|| { \alpha_2 } ||}^2 )} q^{ - \frac{1}{2} \kappa_\alpha  } 
P_{ \alpha_2 ^t } (q^{-\rho};t,q)
%\sum_\sigma  
{ {\left( {\frac{t}{q}} \right)}^{\frac{1}{2}(| \sigma | + | {\beta_2 } | - | \alpha  |)} 
s_{\beta_2 ^t /\sigma } ( q^{ - \rho } t^{ - \alpha_2 } ) s_{ \alpha^t /\sigma } (t^{ - \rho } q^{ - \alpha_2 ^t } )} .
\nonumber
\end{align}
Then one can calculate the sum over $\alpha$ by using (\ref{sum4}) as follows
\begin{align}
&\sum_\alpha  { {\left( { - Q} \right)}^{ - \left| \alpha  \right|}  s_{\alpha /\tau } ( q^{ - \rho }  t^{ - \beta_1^t } ) s_{ \alpha ^t /\sigma } (t^{ - \rho } q^{ -  \alpha _2 ^t } )} \nonumber\\
&=  {( - Q)}^{\left| \tau  \right| + \left| \sigma  \right| - \left| \alpha  \right|} \prod_{i,j = 1}^\infty  {(1 - Q t^{ - \beta_{1,i}^t  + j - \frac{1}{2}} q^{ -  \alpha_{2,j}^t  + i - \frac{1}{2}} )} 
\sum_\alpha  { s_{ \sigma^t /\alpha } (t^{ - \beta_1^t }  q^{ - \rho } ) s_{ \tau^t /  \alpha^t } ( t^{ - \rho }  q^{ -  \alpha_2^t } )}. \nonumber
\end{align}
Thus this partition function becomes
\begin{align}
Z^{(\textrm{a})} (Q,t,q) &= 
{\left( {\frac{q}{t}} \right)}^{\frac{1}{2}( {|\alpha_1| -{|| \alpha_2 ||}^2  + 
{|| {\beta_1 } ||}^2 -|\beta_2| )}}
P_{{\beta_1}^t } (t^{-\rho};q,t)
P_{ {\alpha_2}^t } (q^{-\rho};t,q)
\prod_{i,j = 1}^\infty  {(1 - Q t^{ - \beta_{1,i}^t  + j - \frac{1}{2}} q^{ -  \alpha_{2,j}^t  + i - \frac{1}{2}} )} \nonumber\\
&\times \sum_{\alpha,\sigma,\tau} (-Q)^{|\tau|+|\sigma|-|\alpha|}
\left(\frac{q}{t} \right)^{\frac{|\tau|}{2}-\frac{|\sigma|}{2}}
s_{ \alpha_1 ^t /\tau } ( t^{ - \rho } q^{ -  \beta_1 } ) 
s_{ \tau^t /  \alpha^t } ( t^{ - \rho }  q^{ -  \alpha_2^t } )
s_{\beta_2 ^t /\sigma } ( q^{ - \rho } t^{ - \alpha_2 } )
s_{ \sigma^t /\alpha } (t^{ - \beta_1^t }  q^{ - \rho } ) .\nonumber
\end{align}
Let us denote the second line of the above equation by $\tilde{Z}^{(\textrm{a})}(Q;t,q)$.

A similar calculation gives the partition function for Fig.\ref{floptr}(b)
\begin{align}
Z^{(\textrm{b})} (Q,t,q) &= \sum_{\alpha} (-Q)^{|\alpha|} C_{\alpha, \beta_2, \beta_1}(t,q)  C_{\alpha^t, \alpha_1, \alpha_2} (q,t)\nonumber\\
&= \sum_{\alpha,\tau,\sigma} (-Q)^{|\alpha|}
{\left( {\frac{q}{t}} \right)}^{\frac{1}{2}( {||  \beta_2 ||}^2  + 
{||  {\beta_1 } ||}^2 )} t^{\frac{1}{2} \kappa_{\beta_2}  } 
P_{\beta_1^t } (t^{-\rho};q,t)
%\sum_\tau  
{ {\left( {\frac{q}{t}} \right)}^{\frac{1}{2}(\left| \tau  \right| + \left| { \alpha } \right| - | \beta_2  |)}  s_{ \alpha ^t /\tau } ( t^{ - \rho } q^{ -  \beta_1 } ) s_{\beta_2 /\tau } (q^{ - \rho }  t^{ -  \beta_1 ^t } )} \nonumber\\
&\times {\left( {\frac{t}{q}} \right)}^{\frac{1}{2}( {|| {\alpha_1 } ||}^2
+ {|| { \alpha_2 } ||}^2 )} q^{ \frac{1}{2} \kappa_{\alpha_1}  } 
P_{ \alpha_2 ^t } (q^{-\rho};t,q)
%\sum_\sigma  
{ {\left( {\frac{t}{q}} \right)}^{\frac{1}{2}(| \sigma | + | {\alpha} | - | \alpha_1  |)} 
s_{\alpha /\sigma } ( q^{ - \rho } t^{ - \alpha_2 } ) s_{ \alpha_1 /\sigma } (t^{ - \rho } q^{ - \alpha_2 ^t } )}. \nonumber
\end{align}
After some algebras we get the following expression
\begin{align}
Z^{(\textrm{b})} (Q,t,q) &=
{\left( {\frac{q}{t}} \right)}^{
\frac{1}{2}( 
|\alpha_1|-|\beta_2|
-{|| {\alpha_1 } ||}^2
- {|| { \alpha_2 } ||}^2 
+{||  \beta_1 ||}^2  + 
{||  {\beta_2 } ||}^2 )}
 q^{ \frac{1}{2} \kappa_{\alpha_1}  }  t^{\frac{1}{2} \kappa_{\beta_2}  } 
P_{ \alpha_2 ^t } (q^{-\rho};t,q) P_{\beta_1^t } (t^{-\rho};q,t)\nonumber\\
&\times\prod_{i,j = 1}^\infty  {(1 - Q t^{ - \alpha_{2,i}  + j - \frac{1}{2}} q^{ -  \beta_{1,j}  + i - \frac{1}{2}} )} \nonumber\\
&\times
\sum_{\alpha,\sigma,\tau} (-Q)^{|\tau|+|\sigma|-|\alpha|}
\left(\frac{q}{t} \right)^{\frac{|\tau|}{2}-\frac{|\sigma|}{2}}
s_{ \sigma ^t /\alpha^t } ( t^{ - \rho } q^{ -  \beta_1 } ) 
s_{ \alpha_1 /  \sigma } ( t^{ - \rho }  q^{ -  \alpha_2^t } )
s_{\tau^t /\alpha } ( q^{ - \rho } t^{ - \alpha_2 } )
s_{ \beta_2 /\tau } (t^{ - \beta_1^t }  q^{ - \rho } ). \nonumber
\end{align}
Let us rewrite these partition functions using the free fermions. See the next subsection for the definitions and basic properties of the free fermions. The property (\ref{skewf}) which will be studied in the next subsection enables us to  change the representations in the Schur functions into their transposes. Then the partition function $\tilde{Z}^{(\textrm{a})}$ becomes\begin{align}
\tilde{Z}^{(\textrm{a})}(Q;t,q)&=\sum_{\alpha,\sigma,\tau} (-Q)^{|\tau|+|\sigma|-|\alpha|}
\left(\frac{q}{t} \right)^{\frac{|\tau|}{2}-\frac{|\sigma|}{2}}
s_{ \alpha_1 ^t /\tau } ( t^{ - \rho } q^{ -  \beta_1 } ) 
s_{ \tau^t /  \alpha^t } ( t^{ - \rho }  q^{ -  \alpha_2^t } )
s_{\beta_2 ^t /\sigma } ( q^{ - \rho } t^{ - \alpha_2 } )
s_{ \sigma^t /\alpha } (t^{ - \beta_1^t }  q^{ - \rho } ) \nonumber\\
&=\sum_{\alpha,\sigma,\tau} (-Q)^{|\tau|+|\sigma|-|\alpha|}
\left(\frac{q}{t} \right)^{\frac{|\tau|}{2}-\frac{|\sigma|}{2}} \nonumber\\
&\times
\langle\alpha_1|\omega V_- (t^{-\rho}q^{-\beta_1})|\tau^t\rangle
\langle\tau^t|V_- (t^{ - \rho }  q^{ -  \alpha_2^t })|\alpha^t\rangle
\langle\alpha^t|\omega V_+ (t^{ - \beta_1^t }  q^{ - \rho })|\sigma\rangle
\langle\sigma|V_+ (q^{ - \rho } t^{ - \alpha_2 } )|\beta_2^t\rangle. \nonumber
\end{align}
Using $Q^{L_0}|R\rangle =Q^{|R|}|R\rangle $ and $1=\sum_{\mu}|\mu \rangle\langle \mu|$ we can express it as a single correlation function as
\begin{align}
\tilde{Z}^{(\textrm{a})}(Q;t,q)&=\langle\alpha_1|\omega V_- (t^{-\rho}q^{-\beta_1})\left(-Q\left(\frac{q}{t} \right)^{\frac{1}{2}}\right)^{L_0}
V_- (t^{ - \rho }  q^{ -  \alpha_2^t }) \nonumber\\
&\times(-Q)^{-L_0}
\omega V_+ (t^{ - \beta_1^t }  q^{ - \rho })\left(-Q\left(\frac{q}{t} \right)^{-\frac{1}{2}}\right)^{L_0}
V_+ (q^{ - \rho } t^{ - \alpha_2 } )|\beta_2^t\rangle. \nonumber
\end{align}
Similarly, $\tilde{Z}^{(\textrm{b})}$ becomes
\begin{align}
\tilde{Z}^{(\textrm{b})}(Q;t,q)&=
\sum_{\alpha,\sigma,\tau} (-Q)^{|\tau|+|\sigma|-|\alpha|}
\left(\frac{q}{t} \right)^{\frac{|\tau|}{2}-\frac{|\sigma|}{2}}
s_{ \sigma ^t /\alpha^t } ( t^{ - \rho } q^{ -  \beta_1 } ) 
s_{ \alpha_1 /  \sigma } ( t^{ - \rho }  q^{ -  \alpha_2^t } )\nonumber\\
&\quad\quad\quad\quad\quad\quad\quad\quad\quad\quad\quad\quad\quad\quad\quad\quad\times
s_{\tau^t /\alpha } ( q^{ - \rho } t^{ - \alpha_2 } )
s_{ \beta_2 /\tau } (t^{ - \beta_1^t }  q^{ - \rho } )\nonumber\\
&=
\langle\alpha_1|V_- (t^{ - \rho }  q^{ -  \alpha_2^t })\left(-Q\left(\frac{q}{t} \right)^{-\frac{1}{2}}\right)^{L_0}
\omega V_- (t^{-\rho}q^{-\beta_1})\nonumber\\
&\quad\quad\quad\quad\quad\times (-Q)^{-L_0}
V_+ (q^{ - \rho } t^{ - \alpha_2 } )
\left(-Q\left(\frac{q}{t} \right)^{\frac{1}{2}}\right)^{L_0}
\omega V_+ (t^{ - \beta_1^t }  q^{ - \rho })|\beta_2^t\rangle. \nonumber
\end{align}
This is very similar to $\tilde{Z}^{(\textrm{a})}$. So let us study the relationship between them.
The energy operator acts on the vertex operators as $Q^{L_0}V_{\pm}(x_i)Q^{-L_0} = V_{\pm}(Q^{\mp}x_i)$ and its action commutes with $\omega$-action. From these facts we obtain
\begin{align}
\label{2.1}
\tilde{Z}^{(\textrm{a})}(Q;t,q) = (-Q)^{|\alpha_1|+|\beta_2|}\left(\frac{q}{t} \right)^{\frac{1}{2}(|\alpha_1|-|\beta_2|)} \tilde{Z}^{(\textrm{b})}(Q^{-1};t,q).
\end{align}
One can show the following identity using (3.11) of \cite{Taki:2007dh}
\begin{align}
\label{2.2}
&\prod_{i,j = 1}^\infty  \frac{{(1 - Q t^{  j - \frac{1}{2}} q^{ i - \frac{1}{2}} )}}
{{(1 - Q t^{ - \beta_{1,i}^t  + j - \frac{1}{2}} q^{ -  \alpha_{2,j}^t  + i - \frac{1}{2}} )}} 
\nonumber\\
&=
(-Q)^{-|\alpha_2|-|\beta_1|}
t^{\frac{1}{2}(-||\alpha_2||^2+||\beta_1^t||^2)}
q^{\frac{1}{2}(||\alpha_2^t||^2-||\beta_1||^2)}
\prod_{i,j = 1}^\infty  \frac{{(1 - Q^{-1} t^{  j - \frac{1}{2}} q^{ i - \frac{1}{2}} )}}
{{(1 - Q^{-1} t^{ - \alpha_{2,i}  + j - \frac{1}{2}} q^{ -  \beta_{1,j}  + i - \frac{1}{2}} )}}.
\end{align}
Combining (\ref{2.1}) and (\ref{2.2}), we get the flop invariance of the refined partition function
\begin{align}
\label{flop}
\hat{Z}^{(\textrm{a})} (Q;t,q)=A_{\alpha_1,\alpha_2,\beta_1,\beta_2}(Q;t,q)\hat{Z}^{(\textrm{b})} (Q^{-1}:t,q).
\end{align}
The coefficient $A$ and $\hat{Z}^{(\textrm{n})}$ are such that
\begin{align}
A_{\alpha_1,\alpha_2,\beta_1,\beta_2}(Q;t,q)&= (-Q)^{|\alpha_1|+|\alpha_2|+|\beta_1|+|\beta_2|}
\left(\frac{q}{t}\right)^{\frac{1}{2}(|\alpha_1|-|\beta_2|)}\nonumber\\
&\times
q^{\frac{1}{2}( {|| {\alpha_1^t } ||}^2
- {|| { \alpha_2^t } ||}^2 
+{||  \beta_1 ||}^2
 -{||  {\beta_2 } ||}^2 )}
t^{\frac{1}{2}( -{|| {\alpha_1 } ||}^2
+ {|| { \alpha_2 } ||}^2 
-{||  \beta_1^t ||}^2  + 
{||  {\beta_2^t } ||}^2 )},
\end{align}
\begin{align}
\hat{Z}^{(\textrm{n})}(Q;t,q)=Z^{(\textrm{n})}(Q;t,q)/\prod_{i,j=1}(1-Qt^{i-1/2}q^{j-1/2})
,\quad \textrm{n}=\textrm{a},\textrm{b}.
\end{align}
We can apply the equation (\ref{flop}) for sub-diagrams of refined partition functions. Therefore, we can obtain the flop invariance of the refined topological vertex\footnote[2]{The flop invariance of the refined vertex which was proposed in \cite{Awata:2005fa} is studied in \cite{Awata:2008} } as the result of (\ref{flop}).

\subsection{Proof of formula}
 
 In this section we introduce the free fermion techniques  \cite{Eguchi:2003it} in order to prove a formula for a special class of the Schur functions. The formula plays an important role in the refined topological vertex calculations of the flop transition.
 
 The free fermions in two dimensions are characterized by the anti-commutation relations
\begin{eqnarray}
\left\{\psi_i ,\psi^{*}_j \right\} = \delta_{i+j,0}, \quad \left\{\psi_i ,\psi_j \right\} =\left\{\psi_i^{*} ,\psi^{*}_j \right\} =0,\quad i,j\in\mathbb{Z}+1/2.
\end{eqnarray}
The bosonisation of these fermions gives rise to chiral bosons and their modes are given by
\begin{eqnarray}
\alpha_n = \sum_{j\in\mathbb{Z}+ 1/2} :\psi_{-j+n} \psi^{*}_j :.
\end{eqnarray}
The Young diagrams are in one-to-one correspondence with the elements of the fermionic Fock space such that  
\begin{eqnarray}
 |R\rangle = (-1)^{r(r-1)/2 + \sum_{j=1}^{r(R)}b_j} \prod_{j=1}^{r(R)}{\psi_{-a_j-1/2}  \psi^{*}_{-b_j-1/2}}|0\rangle ,
\end{eqnarray}
where $r(R)$ is a diagonal length of the diagram $R$ and $|0\rangle$ is the Dirac vacuum which is annihilated by $\psi_{i},\psi^{*}_{j}$ for $i,j >0$. The Frobenius coordinates of the Young diagram $a_i$ and $b_i$ are given by
\begin{eqnarray}
a_i (R) = R_i - i, \quad b_i(R) = R^t_i - i.
\end{eqnarray}
The vertex operators which act on these states are defined by
\begin{eqnarray}
V_{\pm}(x_i) = \exp \left[\sum_{i=1,n=1}^{} \frac{x_i^n}{n} \alpha_{\pm} \right].
\end{eqnarray}
The property $\prod_i V_{-}(x_i) |R\rangle = \sum_{Q \supset R}s_{Q/R}(x_i) |Q\rangle$ \cite{kac} allows us to find a free fermion representation of the skew Schur functions
\begin{align}
\label{skew}
s_{R/Q} (x_i) = \langle R|V_{-}(x_i)|Q\rangle =\langle Q|V_{+}(x_i)|R\rangle.
\end{align}

Let us prove the following proposition using the free fermions. The formula for the special case $t=q$ can be found in \cite{macdonald}\cite{Okounkov:2003sp}\cite{Zhou:2003}. 

{\bf Proposition.}
\begin{eqnarray}
\label{prop}
 s_{P/Q} \left(q^{\rho}t^{R} \right) =\omega_{t,q} \cdot s_{P^t / Q^t}  \left( -t^{-\rho}q^{-R^t} \right) .
\end{eqnarray}

Here $\omega_{t,q}$ is an operation act on a power sum\footnote[3]{Notice that the Schur functions are related to the Newton polynomial as 
\begin{align}
s_R (x) = \sum_{\vec{k}} \frac{\chi_R (C(\vec{k}))}{z_{\vec{k}}} P_{\vec{k}}(x), \quad
P_{\vec{k}}(x) = \prod_{n=1}{} p_n^{k_n}(x). \nonumber
\end{align}
} as
\begin{align}
\omega_{t,q} \cdot p_n (x_i) = \frac{[n]_t}{[n]_q} p_n (x_i).
\end{align}

\textbf{Proof:} 
First we introduce an automorphism $\omega$ of the Fock space
\begin{align}
\omega(\psi_j) 
= (-1)^{j} \psi^{*}_j, \quad \omega(\psi^{*}_j) = (-1)^{j} \psi_j.
\end{align}
One can show the following properties
\begin{align}
\omega(|R\rangle) 
= (-1)^{r(r-1)/2 + \sum_{j=1}^{r(R)}a_j + r} \prod_{j=1}^{r(R)}{\psi^{*}_{-a_j-1/2}  \psi_{-b_j-1/2}}|0\rangle ,
=|R^t\rangle.
\end{align}
\begin{align}
\omega(\alpha_n) = (-1)^{n+1} \alpha_n .
\end{align}
Using this automorphism, one can also show that the skew Schur functions (\ref{skew}) become
\begin{align}
\label{skewf}
s_{P/Q} (x) = \langle P^t|\omega(V_{-}(x_i))|Q^t\rangle
= \langle P^t|\exp \left(\sum_{n=1} \frac{1}{n}(-1)^{n+1} p_n (x_i) \alpha_{-n} \right)|Q^t\rangle .
\end{align}
For a demonstration let us prove the well-known formula $s_{P/Q}(q^{-\rho - R}) = (-1)^{|P|-|Q|} s_{P^t / Q^t} 
(q^{\rho + R^t})$ by applying (\ref{skewf}). As we will see the free fermion techniques will simplify the proof of \cite{Zhou:2003}. The power sum of the variables $x_i = q^{i-1/2-R_i}$ is given by
\begin{align}
(-1)^n p_n (x_i = - q^{-i+1/2+R_i}) &= \sum_{i=1}^{\infty} q^{n(-i+1/2+R_i)} \nonumber\\
&= \sum_{i=1}^{\infty} q^{n(-i+1/2)} + \sum_{i=1}^{d(R)} q^{n(-i+1/2)} (q^{nR_i} -1) \nonumber\\
&=1/[n]_q + [n]_q \sum_{i=1}^{d(R)}\sum_{j=1}^{R_i} q^{-ni} q^{nj} , \nonumber
\end{align}
where $[n]_q=q^{n/2}-q^{-n/2}$ is a $q$-number.
Rewriting the last line of the above equation we obtain the relation between the power summations of $q^{\rho + R}$
\begin{align}
(-1)^n p_n (x_i = - q^{-i+1/2+R_i}) 
&=-1/[n]_{q^{-1}} - [n]_{q^{-1}} \sum_{j=1}^{d(R^t)}\sum_{i=1}^{R^t_j} (q^{-1})^{-nj} (q^{-1})^{ni} \nonumber\\
&= -p_n (x_i = -q^{i-1/2-R^t_i}).
\end{align}
Then using $(-1)^{n+1} p_n (q^{-\rho-R}) = p_n (-q^{\rho+R^t})$ and (\ref{skewf}), we get the formula
\begin{align}
s_{P/Q}(q^{-\rho - R}) = \langle P^t|\exp (\sum_{n=1} p_n (-q^{\rho+R^t}) \alpha_{-n} )|Q^t\rangle
=s_{P^t / Q^t} (-q^{\rho + R^t}).
\end{align}

Now we are ready to study the Schur function for the two-parameter variables $x=q^{\rho}t^{R}$. The power sum of these variables becomes
\begin{align}
(-1)^n p_n (x_i = - q^{-i+1/2}t^{R_i}) 
%&= \sum_{i=1}^{\infty} {q^{n(-i+1/2)}}{t^{nR_i}} \nonumber\\
&=1/[n]_t + q^{n/2}t^{-n/2}[n]_t \sum_{i=1}^{d(R)}\sum_{j=1}^{R_i} q^{-ni} t^{nj} \nonumber\\
&=-1/[n]_{t^{-1}} - q^{-n/2}t^{n/2}[n]_{t^{-1}} \sum_{j=1}^{d(R^t)}\sum_{i=1}^{R^t_j} {t^{-1}}^{-nj} {q^{-1}}^{-ni} \nonumber\\
&= -[n]_t/[n]_q p_n (x_i = t^{i-1/2}q^{-R^t_i}) .
\end{align}
Thus we obtain 
\begin{align}
\label{22}
(-1)^{n+1}p_n(q^{\rho}t^{R}) = \omega_{t,q} \cdot p_n (-t^{-\rho}q^{-R^t}),
\end{align}
and the proposition (\ref{prop}) which we are trying to prove follows directly from  (\ref{skewf}) and (\ref{22}).

\section{Flop, Slicing Invariance, and Homological Link Invariants}

\subsection{Homological Link Invariants}
The latest developments in the homological link invariants provide many insights into a refinement of the polynomial invariants. In the theory of homological link invariants, one regards the link invariants as Euler characteristics of graded homologies associated with links $L$
\begin{align}
\bar{\mathcal{P}}^{sl(N)}_{R_1,\cdots, R_k}(\textbf{q})
=\sum_{i,j\in \mathbb{Z}}(-1)^j\textbf{q}^i\dim \mathcal{H}^{sl(N),R_1,\cdots, R_k}_{i,j}(L).
\end{align}
Then we can introduce the homological link invariants as Poincar\'{e} polynomials
\begin{align}
\bar{\mathcal{P}}^{sl(N)}_{R_1,\cdots, R_k}(\textbf{q},\textbf{t})
=\sum_{i,j\in \mathbb{Z}}\textbf{q}^i\textbf{t}^j\dim \mathcal{H}^{sl(N),R_1,\cdots, R_k}_{i,j}(L).
\end{align}
These invariants provide a two-parameter refinement of link invariants.
 The homological link invariants can be embedded into open topological strings \cite{Gukov:2004hz} using an idea of the geometric transition and the refinement of the polynomial invariants is interpreted by using the refined topological vertex. Thus a superpolynomial of the Hopf link which gives the $sl(N)$ homological link invariants for $\textbf{a}=\textbf{q}^N$ has been proposed in \cite{Gukov:2007tf}
\begin{align}
\label{super}
\bar{\mathcal{P}}_{\lambda.\mu}(\textbf{q},\textbf{t},\textbf{a})
&=\sum_{\nu}(-Q)^{|\nu|}t^{\frac{1}{2}||\nu||^2}q^{\frac{1}{2}||\nu^t||^2}
\tilde{Z}_{\nu}(q,t)\tilde{Z}_{\nu^t}(t,q)
s_{\lambda}(t^{-\rho}q^{-\nu^t})s_{\mu}(t^{-\rho}q^{-\nu^t})\nonumber\\
&\quad\quad\times \prod_{i,j=1}(1-Qt^{1-1/2}q^{j-1/2})^{-1}
(-1)^{|\lambda|+|\mu|}
{\left( Q^{-1}\sqrt{\frac{q}{t}} \right)}^{\frac{|\lambda|+|\mu|}{2}}
{\left( \frac{q}{t} \right)}^{|\lambda||\mu|},
\end{align}
where we introduce new variables as
\begin{align}
\label{param}
\sqrt{t}=\textbf{q}, \quad \sqrt{q}=-\textbf{t}\textbf{q},\quad Q=-\textbf{t}/\textbf{a}^2.
\end{align}
The superpolynomials for many representations were calculated in \cite{Gukov:2007tf}. For instance 
\begin{align}
\label{sup1}
\bar{\mathcal{P}}_{\fund,\phi}(\textbf{q},\textbf{t},\textbf{a})
=\frac{\textbf{a}-\textbf{a}^{-1}}{\textbf{q}-\textbf{q}^{-1}},
\end{align}
\begin{align}
\label{sup2}
\bar{\mathcal{P}}_{\anti,\phi}(\textbf{q},\textbf{t},\textbf{a})
=\frac{\textbf{a}^2\textbf{q}^4}{(1-\textbf{q}^2)(1-\textbf{q}^4)}
-\frac{\textbf{q}^4}{(1-\textbf{q}^2)^2}
+\frac{\textbf{a}^{-2}\textbf{q}^6}{(1-\textbf{q}^2)(1-\textbf{q}^4)},
\end{align}
\begin{align}
\label{sup3}
\bar{\mathcal{P}}_{\fund,\fund}(\textbf{q},\textbf{t},\textbf{a})
=\frac{1}{\textbf{a}^2}\left(\frac{1-\textbf{q}^2+\textbf{q}^4\textbf{t}^2}{(1-\textbf{q}^2)^2}
-\textbf{a}^2\frac{1+\textbf{q}^2\textbf{t}^2-\textbf{q}^2+\textbf{q}^4\textbf{t}^2}{(1-\textbf{q}^2)^2}
+\textbf{a}^4\frac{\textbf{q}^2\textbf{t}^2}{(1-\textbf{q}^2)^2}\right).
\end{align}
As discussed in \cite{Gukov:2007tf} the superpolynomial for $(\fundl,\fundl)$ provides the Khovanov-Rozansky homological invariant of the Hopf link $KhR(2_1^2)$
\begin{align}
\bar{\mathcal{P}}_{\fund,\fund}(\textbf{q},\textbf{t},\textbf{a}=\textbf{q}^N)
=\textbf{q}^{-2N}KhR(2_1^2).
\end{align}

\subsection{Homological Link Invariants from Flop}
In the previous section we have calculated the refined partition functions $ Z^{(\textrm{n})}_{(\alpha_1,\alpha_2,\beta_1,\beta_2)}(Q;t,q)$, ($\textrm{n}=\textrm{a,b}$) of Fig.\ref{floptr}. Now we calculate the following normalized partition function
\begin{align}
\label{nom}
\hat{Z}_{\lambda^t,\mu} (Q^{-1};t,q)&\equiv\frac{
 Z^{(\textrm{a})}_{(\lambda^t,\phi,\mu,\phi)}(Q^{-1};t,q)
 }{Z^{(\textrm{a})}_{(\phi,\phi,\phi,\phi)}(Q^{-1};t,q)}
 \nonumber\\
&={\left( \frac{q}{t} \right)}^{\frac{1}{2}(|\lambda|+||\mu||^2)}t^{\frac{1}{2}||\mu||^2}
\tilde{Z}_{\mu}(t,q)
\prod_{i,j = 1}^\infty  \frac{{(1 - Q^{-1}t^{ - \mu_i^t  + j - \frac{1}{2}} q^{ i - \frac{1}{2}} )}} 
{{(1 - Q^{-1} t^{  j - \frac{1}{2}} q^{ i - \frac{1}{2}} )}}
s_{\lambda}\left(t^{-\rho}q^{-\mu}, Q^{-1}\sqrt{\frac{q}{t}} t^{\rho}\right).
\end{align}
Let us calculate the normalized partition function for $(\lambda,\mu)=(\fundl,\phi)$ in the first instance
\begin{align}
\hat{Z}_{\fund,\phi} (Q^{-1};t,q)=-\textbf{t}s_{\fund}(t^{-\rho},\textbf{a}^2t^{\rho})
=-\textbf{t}\left(\frac{\sqrt{t}}{1-t}+\textbf{a}^2\frac{\sqrt{t}}{t-1}\right)
=-\textbf{a}\textbf{t}\frac{\textbf{a}-\textbf{a}^{-1}}{\textbf{t}-\textbf{t}^{-1}}.
\end{align}
Here we have used the new variables (\ref{param}). Thus the partition function can be expressed using the superpolynomial (\ref{sup1}) as
\begin{align}
\label{f1}
\hat{Z}_{\fund,\phi} (Q^{-1};t,q)=-\textbf{a}\textbf{t}\bar{\mathcal{P}}_{\fund,\phi}(\textbf{q},\textbf{t},\textbf{a}). 
\end{align}
Therefore $\hat{Z}_{\fund,\phi} $ is the superpolynomial of the unknot up to an overall factor. We can extend this relation between the conifold partition functions (\ref{nom}) and superpolynomials for some other representations. For $(\lambda,\mu)=(\antis,\phi)$ we have
\begin{align}
\hat{Z}_{\anti^t,\phi} (Q^{-1};t,q)=\textbf{t}^2s_{\anti}(t^{-\rho},\textbf{a}^2t^{\rho}).
\end{align}
Using $s_{\anti}(x)=\frac{1}{2}(\sum_{i,j=1}^{\infty}x_ix_j-\sum_{i=1}^{\infty}x_i^2)$ we obtain
\begin{align}
s_{\anti}(t^{-\rho},\textbf{a}^2t^{\rho})
=\frac{t^2}{(1-t)(1-t^2)}-\frac{\textbf{a}^2}{(1-t)^2}+\frac{\textbf{a}^4t}{(1-t)(1-t^2)}.
\end{align}
This is precisely the superpolynomial for $(\antis,\phi)$ (\ref{sup2}):
\begin{align}
\label{f2}
\hat{Z}_{\anti^t,\phi} (Q^{-1};t,q)=\textbf{a}^2\textbf{t}^2\textbf{q}^{-2}
\bar{\mathcal{P}}_{\anti,\phi}(\textbf{q},\textbf{t},\textbf{a}).
\end{align}
In the last instance we calculate the partition function for $(\fund,\fund)$. Using $Q^{-1}t^{-\frac{1}{2}}q^{\frac{1}{2}}  =\textbf{a}^2$ we have
\begin{align}
\label{3.1}
\hat{Z}_{\fund, \fund} (Q^{-1};t,q)=\frac{q}{t}\sqrt{t}
\frac{1}{1-t}(1-\textbf{a}^2)
s_{\fund}\left(t^{-\rho}q^{-\fund}, \textbf{a}^2t^{\rho}\right).
\end{align}
The Schur function for the single box Young diagram becomes
\begin{align}
s_{\fund}\left(t^{-\rho}q^{-\fund}, \textbf{a}^2t^{\rho}\right)
=\sum_{i=1}^{\infty}(t^{i-1/2}q^{-\delta_{i,1}}+\textbf{a}^2t^{-1+1/2})
=\sqrt{t}\left(\frac{1}{q}-\frac{t}{t-1}+\textbf{a}^2\frac{1}{t-1}\right).
\end{align}
So (\ref{3.1}) gives the superpolynomial (\ref{sup3})
\begin{align}
\label{f3}
\hat{Z}_{\fund,\fund} (Q^{-1};t,q)
&=\frac{1}{(1-t)^2}(1-t+tq+\textbf{a}^2(-1-q+t-tq)+\textbf{a}^4)
\nonumber\\
&=\textbf{a}^2\bar{\mathcal{P}}_{\fund,\fund}(\textbf{q},\textbf{t},\textbf{a}).
\end{align}

From these examples it is very natural to expect that there exists the relation between them such that
\begin{align}
\hat{Z}_{\lambda^t,\mu} (Q^{-1};t,q)=F_{\lambda,\mu}\bar{\mathcal{P}}_{\lambda.\mu}(\textbf{q},\textbf{t},\textbf{a}).
\end{align}
\begin{figure}[tbp]
\begin{center}
\includegraphics[width=11cm,clip]{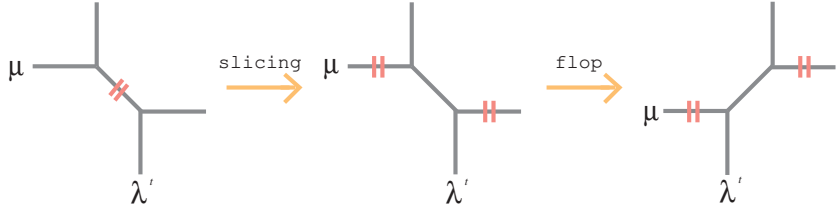}\\
\end{center}
\caption{The superpolynomial $\bar{\mathcal{P}}$ and $\hat{Z}$ are related under the slicing invariance and the flop transition.}
\label{kov}
\end{figure}
Then how can we relate the superpolynomials to the partition functions of Fig.\ref{floptr}(a)?
The slicing invariance proposal \cite{Iqbal:2007ii}\cite{Iqbal:2008ra} plays a key role to convince that the relationship hold water. The slicing invariance means that a refined partition function is independent of the choices of the preferred direction. With this assumption we can relate the partition function $\bar{\mathcal{P}}_{\lambda.\mu}$ to $\hat{Z}_{\lambda^t,\mu}$ as shown in Fig.\ref{kov}. The proportional factor $F_{\lambda,\mu}$ can be determined by comparing the leading terms of their expansions in powers of $Q$. One can expand (\ref{nom}) in powers of $Q$
\begin{align}
\hat{Z}_{\lambda^t,\mu} (Q^{-1};t,q)
={\left( -Q^{-1}\sqrt{\frac{q}{t}} \right)}^{|\lambda|+|\mu|}
{\left( \frac{q}{t} \right)}^{\frac{1}{2}|\lambda|-\frac{1}{2}|\mu|+||\mu||^2}
t^{\frac{1}{2}\kappa_{\lambda}+\frac{1}{2}\kappa_{\mu}+\frac{1}{2}||\mu||^2}  s_{\lambda}(t^{-\rho}) \tilde{Z}_{\mu}(t,q) \times \left(1+ \mathcal{O}(Q)\right) .\nonumber
\end{align}
Here we have used (\ref{2.2}) to change $Q^{-1}$ into $Q$. Similarly, (\ref{super}) becomes
\begin{align}
\bar{\mathcal{P}}_{\lambda.\mu}(\textbf{q},\textbf{t},\textbf{a})=
(-1)^{|\lambda|+|\mu|}{\left( Q^{-1}\sqrt{\frac{q}{t}} \right)}^{\frac{|\lambda|+|\mu|}{2}}
{\left( \frac{q}{t} \right)}^{|\lambda||\mu|}
s_{\lambda}(t^{-\rho})s_{\mu}(t^{-\rho})\times\left(1+ \mathcal{O}(Q)\right) .
\end{align}
Comparing them we expect that the following relationship is satisfied:
\begin{align}
\label{factor}
F_{\lambda,\mu} \equiv\frac{\hat{Z}_{\lambda^t,\mu} (Q^{-1};t,q)}{\bar{\mathcal{P}}_{\lambda.\mu}(\textbf{q},\textbf{t},\textbf{a})}
={\left( Q^{-1}\sqrt{\frac{q}{t}} \right)}^{\frac{|\lambda|+|\mu|}{2}}
{\left( \frac{q}{t} \right)}^{\frac{|\lambda|-|\mu|}{2}-|\lambda||\mu|+||\mu||^2}
 t^{\frac{1}{2}\kappa_{\lambda}+\frac{1}{2}\kappa_{\mu}+\frac{1}{2}||\mu||^2}
\frac{\tilde{Z}_{\mu}(t,q) }{s_{\mu}(t^{-\rho})}.
\end{align}
For instance we can calculate the factor $F_{\lambda,\mu} $ for small representations with easy algebra
\begin{align}
F_{ \fund,\phi} = \textbf{a}(-\textbf{t})\textbf{q}^0= -\textbf{a}\textbf{t}
,\quad
F_{ \anti,\phi}
= \textbf{a}^2\textbf{t}^2 \textbf{q}^{-2}
,\quad
F_{ \fund,\fund} = \textbf{a}^2(-\textbf{t})^0\textbf{q} \frac{1}{\textbf{q} }= \textbf{a}^2.
\end{align}
They agree with our results (\ref{f1}), (\ref{f2}), (\ref{f3}).  Eq.(\ref{factor}) suggests that the following equality is satisfied
\begin{align}
\label{lem2}
&\sum_{\nu}(-Q)^{|\nu|}t^{\frac{1}{2}||\nu||^2}q^{\frac{1}{2}||\nu^t||^2}
\tilde{Z}_{\nu}(q,t)\tilde{Z}_{\nu^t}(t,q)
s_{\lambda}(t^{-\rho}q^{-\nu^t})s_{\mu}(t^{-\rho}q^{-\nu^t})
 \prod_{i,j=1}(1-Qt^{1-1/2}q^{j-1/2})^{-1} \nonumber\\
&={\left(- Q^{-1}\sqrt{\frac{q}{t}} \right)}^{-(|\lambda|+|\mu|)}
{\left( \frac{q}{t} \right)}^{\frac{1}{2}|\mu|-\frac{1}{2}||\mu||^2}
 t^{-\frac{1}{2}(\kappa_{\lambda}+\kappa_{\mu})}\nonumber\\
 &\quad\quad\quad\quad\quad\quad\quad\quad\times
s_{\mu}\left(t^{-\rho}\right)s_{\lambda}\left(t^{-\rho}q^{-\mu}, Q^{-1}\sqrt{\frac{q}{t}} t^{\rho}\right)\prod_{(i,j)\in\mu}(1-Q^{-1}t^{-i+1/2}q^{\mu_i-j+1/2}).
\end{align}
Here we have used the fact $ \prod_{i,j = 1}^\infty  \frac{{(1 - Q^{-1}t^{ - \mu_i^t  + j - \frac{1}{2}} q^{ i - \frac{1}{2}} )}} 
{{(1 - Q^{-1} t^{  j - \frac{1}{2}} q^{ i - \frac{1}{2}} )}}=
\prod_{(i,j)\in\mu}(1-Q^{-1}t^{-i+1/2}q^{\mu_i-j+1/2})$ which has been shown in \cite{Awata:2005fa}. It is an interesting future work to check (\ref{lem2}) using properties of the Macdonald functions.
Therefore we propose the following formula of the superpolynomial of the Hopf link:

{\bf Conjecture.}
\begin{align}
\label{conj}
\bar{\mathcal{P}}_{\lambda,\mu}(\textbf{q},\textbf{t},\textbf{a})&=
{\left( Q^{-1}\sqrt{\frac{q}{t}} \right)}^{-\frac{|\lambda|+|\mu|}{2}}
{\left( \frac{q}{t} \right)}^{\frac{1}{2}|\mu|-\frac{1}{2}||\mu||^2+|\lambda||\mu|}
 t^{-\frac{1}{2}(\kappa_{\lambda}+\kappa_{\mu})}\nonumber\\
 &\quad\quad\times
s_{\mu}\left(t^{-\rho}\right)s_{\lambda}\left(t^{-\rho}q^{-\mu}, Q^{-1}\sqrt{\frac{q}{t}} t^{\rho}\right)\prod_{(i,j)\in\mu}(1-Q^{-1}t^{-i+1/2}q^{\mu_i-j+1/2}).
\end{align}
It is easy to see that (\ref{conj}) reduces to the well known Hopf link invariant if you take $t=q$. However we don't have the rigorous proof. One of difficulties is that we don't have a general mathematical theory of the link homologies. However we can check (\ref{conj}) for some other representations. In the case of the unknot colored by a totally anti-symmetric representation $\antin  =\Lambda^n=\left\{ 1,1,1,\cdots,1\right\}$, it has been observed in \cite{Gukov:2007tf} that the superpolynomial is independent of $\textbf{t}$. Our result (\ref{conj}) implies this property for $\lambda=\Lambda^n$, $\mu=\phi$
\begin{align}
\label{lem}
\bar{\mathcal{P}}_{\Lambda^n,\phi}(\textbf{q},\textbf{t},\textbf{a})&=
\textbf{a}^{-n}
 \textbf{q}^{n(n-1)}
s_{\Lambda^n}\left(\textbf{q}^{-2\rho}, \textbf{a}^2\textbf{q}^{2\rho}\right).
\end{align}
Furthermore, it has been proposed in \cite{Khovanov:2004}\cite{Gukov:2005qp} that the homological link invariant of the unknot $\bar{\mathcal{P}}^{sl(N)}_{\Lambda^n}(\textbf{a}=\textbf{q}^N)=\bar{\mathcal{P}}_{\Lambda^n,\phi}(\textbf{t},\textbf{q})=\sum_{i,j} \textbf{q}^i\textbf{t}^j\dim \mathcal{H}_{i,j}^{sl(N),\Lambda^n}$ is given by the cohomology ring of the Grassmannian $\mathcal{H}_*^{sl(N),\Lambda^n}\simeq H^*(Gr(n,N))$. Therefore (\ref{lem}) should be interpreted in terms of a generating function for the dimensions of this ring. Indeed this expectation is correct. Recall that the Schur function for a totally anti-symmetric representation $\Lambda^n$ is given by
\begin{align}
s_{\Lambda^n}(x)=\sum_{1\leqslant i_1<i_2<\cdots <i_n}x_{i_1}\cdots x_{i_n}.
\end{align}
Therefore the following equation gives a generating function of them
\begin{align}
E(t,\theta)\equiv
\sum_{n}s_{\Lambda^n}\left({t}^{-\rho}, \textbf{a}^2{t}^{\rho}\right)\theta^n
=\prod_{i=1}(1+{t}^{i-1/2}\theta)(1+\textbf{a}^2t^{-i+1/2}\theta).
\end{align}
Using an identity
$
E(t,t\theta)=\frac{1+\textbf{a}^2\sqrt{t}\theta}{1+\sqrt{t}\theta}E(t,\theta),
$
 one can show a recursion relation
\begin{align}
(t^n-1)s_{\Lambda^n}\left({t}^{-\rho}, \textbf{a}^2{t}^{\rho}\right)=
\sqrt{t}(\textbf{a}^2-t^{n-1})s_{\Lambda^{n-1}}\left({t}^{-\rho}, \textbf{a}^2{t}^{\rho}\right).
\end{align}
Thus we obtain
\begin{align}
s_{\Lambda^n}\left({t}^{-\rho}, \textbf{a}^2{t}^{\rho}\right)=
t^{n/2}\frac{(t^{n-1}-\textbf{a}^2)(t^{n-2}-\textbf{a}^2)\cdots(1-\textbf{a}^2)}{(1-t^n)(1-t^{n-1})\cdots (1-t)}.
\end{align}
Substituting it for (\ref{lem}) we get the following expression of the superpolynomial
\begin{align}
\bar{\mathcal{P}}_{\Lambda^n,\phi}(\textbf{q},\textbf{t},\textbf{a})&=
\textbf{a}^{-n}
 \textbf{q}^{n^2}
\frac{(\textbf{q}^{2(n-1)}-\textbf{a}^{2})(\textbf{q}^{2(n-2)}-\textbf{a}^{2})\cdots(1-\textbf{a}^2)}{(1-\textbf{q}^{2n})(1-\textbf{q}^{2(n-1)})\cdots (1-\textbf{q}^2)}.
\end{align}
For $\textbf{a}=\textbf{q}^N$ it gives
\begin{align}
\bar{\mathcal{P}}_{\Lambda^n,\phi}(\textbf{q},\textbf{t},\textbf{a}=\textbf{q}^N)=
\textbf{q}^{2n^2-n(1+N)}
\frac{(1-\textbf{q}^{2})\cdots (1-\textbf{q}^{2(N-1)})(1-\textbf{q}^{2N})}{
(1-\textbf{q}^{2(N-n)})\cdots (1-\textbf{q}^2)\cdot
(1-\textbf{q}^{2n})(1-\textbf{q}^{2(n-1)})\cdots (1-\textbf{q}^2)}\nonumber
\end{align}
This is precisely the Hilbert series of the Grassmannian (see, e.g., \cite{reiner})
\begin{align}
\textrm{Hilb}(H^*(Gr(n,N),\mathbb{Q})))&=\sum_i \dim H^*(Gr(n,N),\mathbb{Q}) _i\textbf{q}^{2i}\nonumber\\
&=\frac{(1-\textbf{q}^{2})\cdots (1-\textbf{q}^{2(N-1)})(1-\textbf{q}^{2N})}{
(1-\textbf{q}^{2(N-n)})\cdots (1-\textbf{q}^2)\cdot
(1-\textbf{q}^{2n})(1-\textbf{q}^{2(n-1)})\cdots (1-\textbf{q}^2)}
\end{align}
Thus our conjecture is consistent with the results of \cite{Khovanov:2004}\cite{Gukov:2005qp}\cite{Gukov:2007tf}.

%%%%%%%%%%%%%%%%%%%%%%%%%%%%%%%%%%%%%%%%%%%%%%%%
\section{Conclusion}
In this article we prove the flop invariance of the reined topological vertex. A technical point is the equation (\ref{skewf}) which enables us to change the representations of the Schur functions into their transposes. Thus we can evaluate the partition functions for Fig.\ref{floptr} and confirm the flop invariance.

Then we compare the partition function of Fig.\ref{floptr}(a) with the $sl(N)$ homological link invariants of the Hopf link. We can see the agreement between them for some small representations. Therefore we propose the expression of the superpolynomial as the conifold partition function. The reason why the relation is satisfied is the slicing invariance of the refined partition function. Once one adopt this assumption one can relate the superpolynomial to the partition function of Fig.\ref{floptr}(a). The new formula is essentially an expression as a product of the Schur functions and therefore  saves some computational costs to get the polynomial expressions of the homological link invariants. This is because if we apply our result, we have no need to compute the sum of the Macdonald functions (\ref{super}) by expanding it in powers of $Q$. Furthermore we expect that our proposal (\ref{conj}) provides some insights into the study of the homological link invariants.
%%%%%%%%%%%%%%%%%%%%%%%%%%%%%%%%%%%%%%%%%%%%%%%%%%%

%%%%%%%%%%%%%%%%%%%%%%%%%%%%%%%%%%%%%%%%%%%%%%%%%%%
\section*{Acknowledgements}
I would like to thank Tohru Eguchi, Hiroaki Kanno and Hiryuki Fuji for valuable discussions and helpful comments. I am thankful to  Hidetoshi Awata and Hiroaki Kanno for sharing their paper\cite{Awata:2008}.

\appendix
\section*{Appendix}
\section{Conventions and Formulas }
\label{appendix:a}

A Young diagram is described as a sequence of decreasing non-negative integers such that
\begin{eqnarray}
\mu  = \left\{ {\mathop \mu \nolimits_i  \in \mathop {\mathbb Z}\nolimits_{ \ge 0} |\mathop \mu \nolimits_1  \ge \mathop \mu \nolimits_2  \ge  \cdots } \right\}.
\end{eqnarray}
The transpose of $\mu $ is defined as follows
\begin{eqnarray}
\mathop \mu \nolimits^t  = \left\{ {\mathop \mu \nolimits_j^t  \in \mathop Z\nolimits_{ \ge 0} |\mathop \mu \nolimits_j^t  = \# \left\{ {i|\mathop \mu \nolimits_i  \ge j} \right\}} \right\}.
\end{eqnarray}
It is useful to define the following quantities
\begin{align}
\left| \mu  \right| = \sum\limits_{i = 1}^{d(\mu )} { \mu_i } ,\quad   \left\| \mu  \right\|^2  = \sum_{i = 1}^{d(\mu )} { \mu_i^2 } 
,\quad {n(\mu ) = \sum_{i = 1}^{d(\mu )} {(i - 1) \mu_i } } ,\quad { \kappa_\mu   = \sum_{(i,j) \in \mu } {(j - i)} }  .
\end{align}
Then one can show the following identities 
\begin{align}
n(\mu ) = \frac{1}{2}\sum_{j = 1}^{ \mu_1 } { \mu_j^t ( \mu_j^t  - 1)},\quad
n(\mu^t ) = \frac{1}{2}\sum_{i = 1}^{d(\mu )} { \mu_i ( \mu_i  - 1)} 
,\quad
{\mathop \kappa \nolimits_\mu   = 2(n(\mathop \mu \nolimits^t  ) - n(\mu)) = \mathop {\left\| \mu \right\|}\nolimits^2  -{\left\| { \mu^t }  \right\|}^2 }.
\end{align}
The following identities are very useful to calculate summations of the Shur functions:
\begin{eqnarray}
\label{sum1}
\sum_\mu  { s_\mu  (x) s_\mu  (y)}  = \prod_{i,j} {(1 -  x_i  y_j )^{ - 1} } ,
\end{eqnarray}
\begin{eqnarray}
\sum_\mu  { s_{\mu ^t } (x) s_\mu  (y)}  = \prod_{i,j} {(1 +  x_i y_j )} ,
\end{eqnarray}
\begin{eqnarray}
\label{sum3}
\sum\limits_\mu  {\mathop s\nolimits_{\mu /\rho } (x)\mathop s\nolimits_{\mu /\sigma } (y)}  = \prod\limits_{i,j} {\mathop {(1 - \mathop x\nolimits_i \mathop y\nolimits_j )}\nolimits^{ - 1} } \sum\limits_\nu  {\mathop s\nolimits_{\rho /\nu } (y)\mathop s\nolimits_{\sigma /\nu } (x)} ,
\end{eqnarray}
\begin{eqnarray}
\label{sum4}
\sum_\mu  { s_{\mu ^t /\rho } (x) s_{\mu /\sigma } (y)}  = \prod_{i,j} {(1 +  x_i  y_j )} \sum_\nu  { s_{\rho /\nu ^t } (y) s_{\sigma ^t /\nu ^t } (x)} ,
\end{eqnarray}
\begin{eqnarray}
s_\mu  (Qx) =  Q^{\left| \mu  \right|} s_\mu  (x),
\end{eqnarray}
\begin{eqnarray}
 s_{\mu /\nu } (Qx) = Q^{\left| \mu  \right| - \left| \nu  \right|}  s_{\mu /\nu } (x),
\end{eqnarray}
\begin{eqnarray}
s_\mu  ( q^\rho  ) =q^{\frac{{\kappa_\mu  }}{2}} s_{\mu^t } (q^\rho  ) ={( - 1)}^{\left| \mu  \right|} s_{ \mu ^t } ( q^{ - \rho } ).
\end{eqnarray}
%%%%%%%%%%%%%%%%%%%%%%%%%%%%%%%%%%%%%%%%%%%%%%%%%%%

\end{document}